**Stabilization and functional properties of $La_3NiAlMnO_9$ and $La_3CoAlMnO_9$ magnetoelectric triple perovskites**

**M.P. Singh \*, K.D. Truong, S. Jandl, and P. Fournier**

Département de Physique and Regroupement québécois sur les matériaux de pointe, Université de Sherbrooke, Sherbrooke (QC), J1K 2R1 Canada

## Abstract

Ferromagnetic $La_3NiAlMnO_9$ (LNAMO) and $La_3CoAlMnO_9$ (LCAMO) triple-perovskite thin films are stabilized in the 750-860 $^oC$ temperature range in 100 to 900 mTorr $O_2$ pressure range using pulsed-laser deposition. The LCAMO and LNAMO films exhibit ferromagnetism up to 190 K and 130 K respectively. The structural, optical and magnetic properties of these films demonstrate that the B-site 3d-cations, Al, Mn and Co or Ni ions, are structurally short-range ordered. The strong spin-lattice-polarization coupling in LCAMO is evidenced by the temperature dependence of the dielectric constant and the softening of the phonon frequencies starting in the vicinity of the ferromagnetic-to-paramagnetic phase transition mimicking the behaviours of $La_2CoMnO_6$ double perovskite.

\* Corresponding author (Email: mangala.singh@usherbrooke.ca)

Well controlled magnetodielectric response in materials is of great importance from both technological and fundamental points of view[1-2]. This has driven the current ongoing search for alternative strategies to design innovative multiferroics. The common approaches rely on mixing magnetism and the dielectric response either in the form of composites [2], superlattices/multilayers [3] or through doping in appropriate host compounds [1-3]. Self-ordered double perovskites [4] of formula unit $A_2B'MnO_6$ (*e.g.,* B' = Co or Ni; and A = La) represent an alternative avenue to search for such innovative materials [4-15]. They are ferromagnetic insulators driven by superexchange interaction [5] with transition temperatures approaching room temperature. Structural self-ordering of the Mn and B' atoms with different oxidation states obtained in very specific growth conditions leads also to charge ordering similar to that observed in the manganites and a strong local (alternating) crystal electric field with a possible coupling between their polar and magnetic order parameters [4-7]. In our recent studies of self-ordered $La_2CoMnO_6$ (LCMO) thin films grown by pulsed-laser deposition, we demonstrated that the ordered Co/Mn cations can be obtained only in a limited window of growth parameters [11]. It has been clearly established that the magnetic behaviour of these compounds is a very sensitive probe of the presence of cation ordering [5-14]. Furthermore, the $MnO_6$ and $CoO_6$ layers with $Mn^{4+}$ and $Co^{2+}$ ions alternating along the (111) crystal axis generate a strong polar character as confirmed by an enhanced dielectric constant [14] and a large magnetodielectric coupling in the vicinity of the magnetic phase transition [7, 14-15].

In order to learn more on the interplay of the polar character and the magnetic properties and harness their functional properties, it becomes interesting to explore closely-related structures where one can change and tune the strength and nature of the electric and magnetic interactions. In order to modify the coupling between the magnetic and polar properties, we are also searching for innovative routes to break the magnetic symmetry by substituting a non-magnetic ion into the double-perovskite materials while preserving cation ordering and ferromagnetism. To achieve this objective, we present in this paper the effect of partial substitution in equal parts of the B' and Mn sites by non-magnetic Al in the parent compounds $La_2B'MnO_6$ (B'=Co and Ni) leading to the basic (disordered) formula units $LaCo_{0.33}Al_{0.33}Mn_{0.33}O_3$ or $LaNi_{0.33}Al_{0.33}Mn_{0.33}O_3$, respectively.

Ideally, these systems once self-ordered will have unit cell formulas $La_3CoAlMnO_9$ or $La_3NiAlMnO_9$, respectively. As far as we know, these materials have not yet been reported either in bulk or thin film forms. We explore the growth conditions to favour three dimensional (3D) self-ordering of the Al, Co and Mn ions leading to ferromagnetic $La_3CoAlMnO_9$ (LCAMO) and $La_3NiAlMnO_9$ (LNAMO) triple perovskites with evidence of spin-phonon coupling and substantial magnetodielectric response in proximity to their ferromagnetic transition.

The LCAMO and LNAMO triple perovskite thin films on (001) $SrTiO_3$ were grown in the 720-860 °C temperature range by ablating the appropriate stoichiometric target using a KrF excimer laser ($\lambda$ = 248 nm) at 6 Hz. The stoichiometric LCAMO and LNAMO targets were synthesised by standard solid-state synthesis routes. The cationic stoichiometry of our films and targets were examined and confirmed using energy dispersive spectroscopy (EDS) associated with a scanning electron microscope. The best quality epitaxial LCAMO films were grown in the 820-860 °C temperature range under fairly high oxygen pressure of 600 mTorr. Epitaxial LNAMO films were grown around 800 °C under similar high oxygen pressure of 800 mTorr. These optimal conditions approach those for the growth of LCMO [11]. Following the ablation, the deposition chamber was filled with 400 Torr $O_2$ pressure and the samples were cooled down to room temperature at a rate of 10 °C/min.

The epitaxial nature and crystalline quality of the LCAMO and LNAMO films were examined by X-ray diffraction using Cu-K$\alpha$ radiation in the θ-2θ and rocking curve modes. Typical XRD patterns of the LCAMO and LNAMO films grown on $SrTiO_3$ are shown in Fig. 1. The XRD patterns are indexed based on the pseudo-cubic notation. The films are characterized by only one set of (*00l*) reflections (where *l* = 1, 2, 3, etc) demonstrating a *cube-on-cube* relationship with the substrate, its preferred orientation, and most importantly the absence of any impurity phase. To verify the coherent nature of our films, ω-scans (*i.e.*, rocking curves) were also measured on these (*00l*) reflections. A typical rocking curve measured on the (002) reflection for LCAMO films is shown in the inset of Fig. 1a. The full-width at half-maximum (FWHM) of this curve is about 0.7° underlining the high quality of these single-phase films growing coherently. Using the XRD patterns, we estimated the out-of-plane lattice parameter "c" of LCAMO and LNAMO films. Inserting Al in the crystal structure is expected to decrease the lattice parameters of the average primitive cell with respect to LCMO and LNMO. For LCAMO, the out-of-plane pseudo-cubic lattice parameter is about 4Å. For LNAMO, it is about 3.99 Å. Surprisingly, these artificial oxides show a significant enhancement in their pseudo-cubic lattice parameter compared to their respective LCMO (~ 3.89 Å) and LNMO (~ 3.87 Å) bulk and thin films [6-14]. We interpret this counter-intuitive behaviour as an indication that the presence of Al in the crystal structure provokes an expansion of the out-of-plane lattice parameter to minimize the elastic and columbic energy of these polar systems. At a fixed growth temperature, the c-lattice parameter of these films grows with increasing $O_2$ pressure thus showing a trend similar to that of LCMO films [11]. Further work is warranted to study comprehensively the crystal structure and symmetry of these oxides.

The magnetic field dependence of the magnetization (*i.e.* M-H loops) and its temperature dependence (*i.e.,* M-T curves) at a 500 Oe applied magnetic field were measured using a SQUID magnetometer from Quantum Design. Typical in-plane M-H data of LCAMO and LNAMO measured at 10 K (Fig. 2a-b) exhibit well-defined loops with very different coercive fields ($H_c$) of 4.5 kOe and 0.17 kOe, respectively. The saturation magnetization ($M_s$) value, estimated from the M-H loop, is roughly 6.1 $\mu_B$/f.u. for ordered LCAMO and 5.1 $\mu_B$/f.u. for ordered LNAMO. Since $Al^{3+}$ is a non-magnetic ion and that the magnetization purely arises from the Co/Ni and Mn cations, the value found is extremely close to the theoretical expectations of 6 $\mu_B$/f.u. for the magnetic moments of $Co^{2+}$ and $Mn^{4+}$ in their high-spin configurations in LCAMO and 5 $\mu_B$/f.u. for $Ni^{2+}$ and $Mn^{4+}$ in LNAMO [5-6, 10-14]. This illustrates that the LCAMO films include $Co^{2+}$ and $Mn^{4+}$ ions while LNAMO films possess $Ni^{2+}$ and $Mn^{4+}$ ions in the crystal structure (here, Al is assumed to preserve its 3+ oxidation state and carrying no magnetic moment). Any deviation from these chemical states (for example with $Mn^{3+}$ and $Co^{3+}$) would result in lower expectation values for the saturation magnetization as is observed for disordered LCMO and LNMO [5-6, 10-14]. Moreover, even in the presence of ordered $Co^{2+}/Al^{3+}/Mn^{4+}$ ions along specific crystallographic should result in a smaller local electric field in these triple perovskites than in LCMO. Overall, even the coupling between the resulting magnetic and polar orders is likely to decrease.

Stabilizing the complex self-ordered triple-perovskite structure requires very specific growth conditions as was shown for double-perovskite LCMO [9-12]. If the conditions are not set properly, the Al, Mn and B' ions arrange randomly in the structure while preserving the basic $ABO_3$ blocks [4] with unit cell formula $LaB'_{0.33}Al_{0.33}Mn_{0.33}O_3$. In optimized conditions, self-ordering leads to a 3D structure with alternating B', Al and Mn ions in orthogonal directions with a unit cell formula $La_3B'AlMnO_9$. The presence of Al in the Mn-O-Co matrix is expected to influence the physical properties. For example, it imposes anew superexchange path for the $Co^{2+}$-O-$Al^{3+}$-O-$Mn^{4+}$ bonds in the ordered system and a $Co^{3+}$-O-$Al^{3+}$-O-$Mn^{3+}$, $Al^{3+}$-O-$Co^{3+}$-O-$Mn^{3+}$ bond mixture in the disordered phase. In both cases, the presence of Al should reduce the strength of the magnetic superexchange interaction [5] and alter significantly their polar behaviour in proximity to the ferromagnetic transition. Consequently, we are expecting a decrease in the ferromagnetic transition Curie temperature (FM-$T_c$) with respect to LCMO and LNMO.

The M-T curves (insets of Figs. 2a and b) for LCAMO and LNAMO clearly show that the magnetization is independent of temperature at low temperature and is presenting a ferromagnetic behavior up to 190 K and 130 K respectively. These values of FM-$T_c$ are significantly lower than the parent phases LCMO (~ 250 K) [11] and LNMO (~ 290 K) [7]. The observed difference in FM-$T_c$ of these triple perovskites compared to their parent double-perovskite phases could be understood based on the 180°-superexchange mechanism proposed by Goodenough-Kanamori [5]. The magnetic transition temperature is governed by the magnitude of the spin-transfer integral. Its value is directly determined by the degree of orbital overlap along the 3d-O-3d bonds which varies exponentially with the bond length. In self-ordered LCMO and LNMO, the proximity of Mn and Co/Ni leads to a large overlap. The insertion of $AlO_6$ octahedra in LCMO/LNMO provokes a decrease of the overlap of Mn-O-Co and Mn-O-Ni wave functions which significantly reduces the strength of the superexchange interaction.

Interestingly, the LCAMO films also display a second magnetic transition at low temperature (~ 100 K) marked as FM-Tc$_2$ in Fig. 2a. This low temperature magnetic transition is likely arising from partial cationic disorder as was observed in LNMO [12]. A M-H loop at 125 K (inset of Fig 2a) confirms the persistence of the ferromagnetic behaviour above FM-Tc$_2$ with a small coercive field (200 Oe) and a saturation magnetization of 1.98 µ$_B$/f.u. These magnetic signatures suggest that the films are likely short-range ordered. It may also indicate that larger growth temperatures above 860 °C may be required to remove completely the remaining small proportion of cation-disordered phase. This range of temperature is presently inaccessible with our current PLD system. Finally, our detailed study [16] of the growth conditions for LNAMO and LCAMO shows that the films grown under low oxygen pressures and/or low growth temperatures have a significant amount of this disordered (random) phase.

To study the presence and magnitude of the spin-phonon coupling, we investigated the LCAMO thin films by polarized Raman spectroscopy, which has been a proven technique to investigate local lattice distortions, cation ordering, spin-lattice coupling and phase transitions in ferromagnetic insulators as La$_2$CoMnO$_6$ and La$_2$NiMnO$_6$ epitaxial films [8-9]. We measured the temperature dependence of Raman phonon frequencies from 10 to 300 K using a Labram 800 microscope spectrometer equipped with a He-Ne laser ($\lambda$ = 632.8 nm). The sample was cooled down in a Janis Research Supertran Cryostat and spectra were recorded using a nitrogen cooled charge coupled device (CCD) detector [9]. Raman spectra obtained for different polarization configurations XX, XY, X'Y', X'X' at 300 K are presented in Fig 3a. Several important features can be immediately noticed. *First,* the Raman active phonons and their intensities are very sensitive to the polarization configurations consistent with the crystal symmetry of the films and confirming the epitaxy [7]. *Second*, these phonon peaks are narrow with FWHM of about 42 cm$^{-1}$ at 300K confirming the homogenous and high crystallinity of our samples. *Third,* the number of observed Raman active modes is smaller when compared to self-ordered La$_2$CoMnO$_6$ films with long-range structural order [9]. In particular, we see no evidence of Brillouin-zone folding as was demonstrated for long range ordered LCMO [9]. In fact, the number of phonon modes observed in LCAMO is comparable to that of $La_2NiMnO_6$ that shows only short-range structural order [8]. Taking into account the structural, magnetic, and Raman properties, we conclude that the cations in these triple-perovskite oxides are ordered only on short length scales [8-9].

Despite the absence of long-range order, evidence for spin-lattice coupling in LCAMO can be demonstrated using the temperature dependence of the phonon mode at 661 cm$^{-1}$ originating from the stretching of the (Al/Co/Mn)O$_6$ octahedra as observed in LCMO and LNMO [8-9]. As shown in Figs. 3b and c, this mode starts to soften weakly at the magnetic transition temperature revealing the first glimpse of a magnetoelastic coupling similar to that observed in partially ordered and self-ordered LCMO [8-9]. In a previous report on LCMO, it was shown that the softening of the same mode is reaching a magnitude as large as ~ 10 cm$^{-1}$ from $T_c$ to low temperature (10 K). However in the present case, the total softening reaches barely 3 cm$^{-1}$ [8-9]. This illustrates that the incorporation of AlO$_6$ octahedra in the crystal structure prevents partially the softening of this stretching mode. Furthermore, since this band comprises the stretching phonons arising from the three types of octahedra, one expects also this mode to present a moderate value of its FWHM. The temperature dependence of the FWHM of the 661 cm$^{-1}$ mode is plotted in Fig. 3d showing its decrease to roughly 32 cm$^{-1}$ at low temperature comparable to the observations on short-range ordered films of LCMO and LNMO [8-9].

We also measured the temperature dependence of the dielectric constant of LCAMO films on Nb-doped STO (001). These measurements at 10 kHz using a 1V AC excitation were done using the magnetic field and temperature control of a Physical Properties Measurements System (from Quantum Design) interfaced with external instruments in a simple set-up described elsewhere [14]. At low temperatures, LCAMO is characterized by a dielectric constant (Fig. 4a) value of about ε ~ 11 at 10 K whereas it approaches ε ~ 88 at 250 K. The trend in temperature is quite similar to that of LCMO [14] as the maximum dielectric constant (~ 250) is observed close to the magnetic transition temperature. In fact, the derivative of the dielectric constant as a function of temperature reaches a maximum at 190 K very close to the magnetic $T_c \sim 200$ K. However, the maximum dielectric constant value at the ferromagnetic transition of LCAMO remains lower than the parent LCMO phase [14]. This reduction in the dielectric constant may be attributed to the short-range cationic ordering in our films but also to the fact that the local electric field is expected to be lower as mentioned above.

Finally, we have determined the magnetodielectric (MD) effect by measuring the dielectric constant as a function of temperature in 0, 0.5, 10, 30 kOe applied magnetic fields with the same set-up. The magnetodielectric effect is defined as MD (H) = $\varepsilon$ (H) - $\varepsilon$ (0)/ $\varepsilon$(0). The maximum value for the MD effect (Fig. 4b) is about 0.3% and it vanishes above 270 K. Compared to LCMO, the amplitude of the magnetodielectric effect in LCAMO is significantly suppressed illustrating that the coupling between the polar and magnetic order parameters has been reduced drastically as expected [7, 14-15] due to the incorporation of a non-magnetic octahedra. In summary, we have designed and studied the polar ferromagnetic triple perovskites $La_3CoAlMnO_9$ and $La_3NiAlMnO_9$. Our study demonstrates that the presence of non-magnetic $AlO_6$ octahedra in the $CoO_6/MnO_6$ and $NiO_6/MnO_6$ matrices affects significantly the functional properties when compared to those of $La_2CoMnO_6$ and $La_2NiMnO_6$.

We thank S. Pelletier and M. Castonguay for their technical support. This work was supported by *CIFAR*, *CFI*, *NSERC* (Canada), *FQRNT* (Québec) and the Université de Sherbrooke.

**List of figure captions:**

**Figure 1.** Typical θ-2θ XRD patterns of the **(a)** LCAMO film grown at 850 °C and 600 mTorr $O_2$, and **(b)** LNAMO films grown at 800 °C at 800 mTorr $O_2$. The inset in **(a)** shows a rocking curve recorded from the (002) reflection of LCAMO films. XRD peaks are indexed based on the pseudo-cubic notation.

**Figure 2.** Typical in-plane M-H loop of **(a)** LCAMO films and **(b)** LNAMO films recorded at 10 K. The top-corner inset of Fig. 2a shows the M-H loop of LCAMO films measured at 125K above FM-$Tc_2$ and its right-bottom inset shows the corresponding M-T curve at 500 Oe. Inset of Fig. 2b shows M-T curve LNAMO films at 500 Oe. *Y*-axis of insets represents the value of magnetization in $\mu_B$/f.u.

**Figure 3.** Polarized Raman spectra **(a)** measured at 300 K in different polarization configurations, **(b)** measured as a function of temperature in the XX configuration. Temperature dependence of **(c)** the peak position and **(d)** the full-width-at half maximum for the 661 $cm^{-1}$ phonon mode.

**Figure 4 (a)** The variation in dielectric constant as a function of temperature, and **(b)** the temperature dependence of the dielectric constant measured under different magnetic fields. The inset in **(a)** shows dε/dT as a function of temperature curve displaying the impact of the ferromagnetic transition on the dielectric constant.

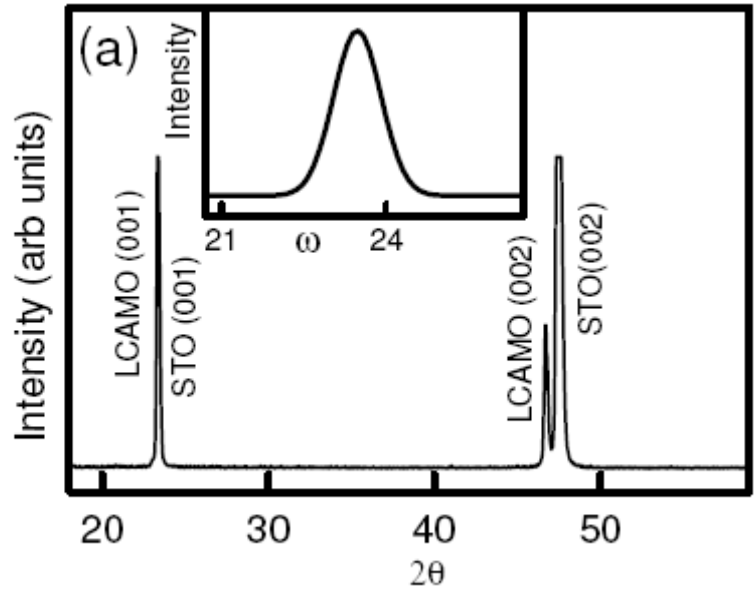

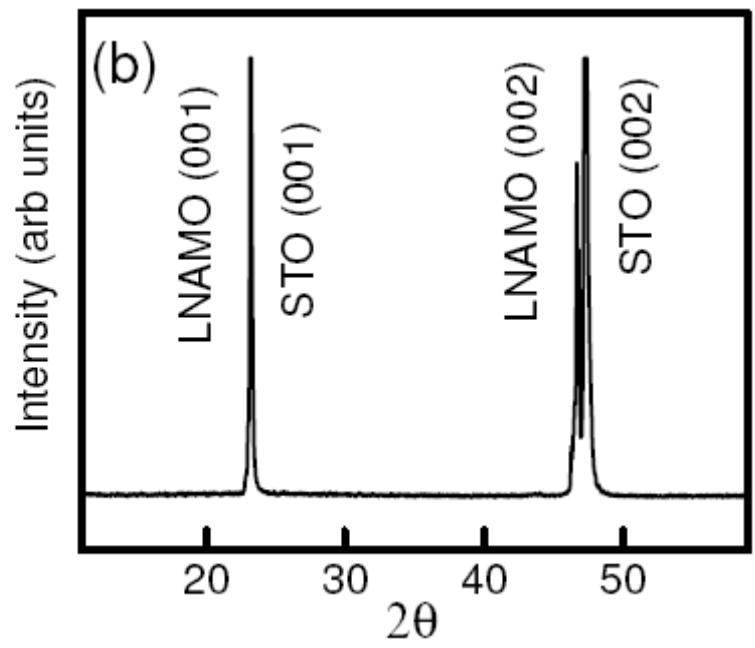

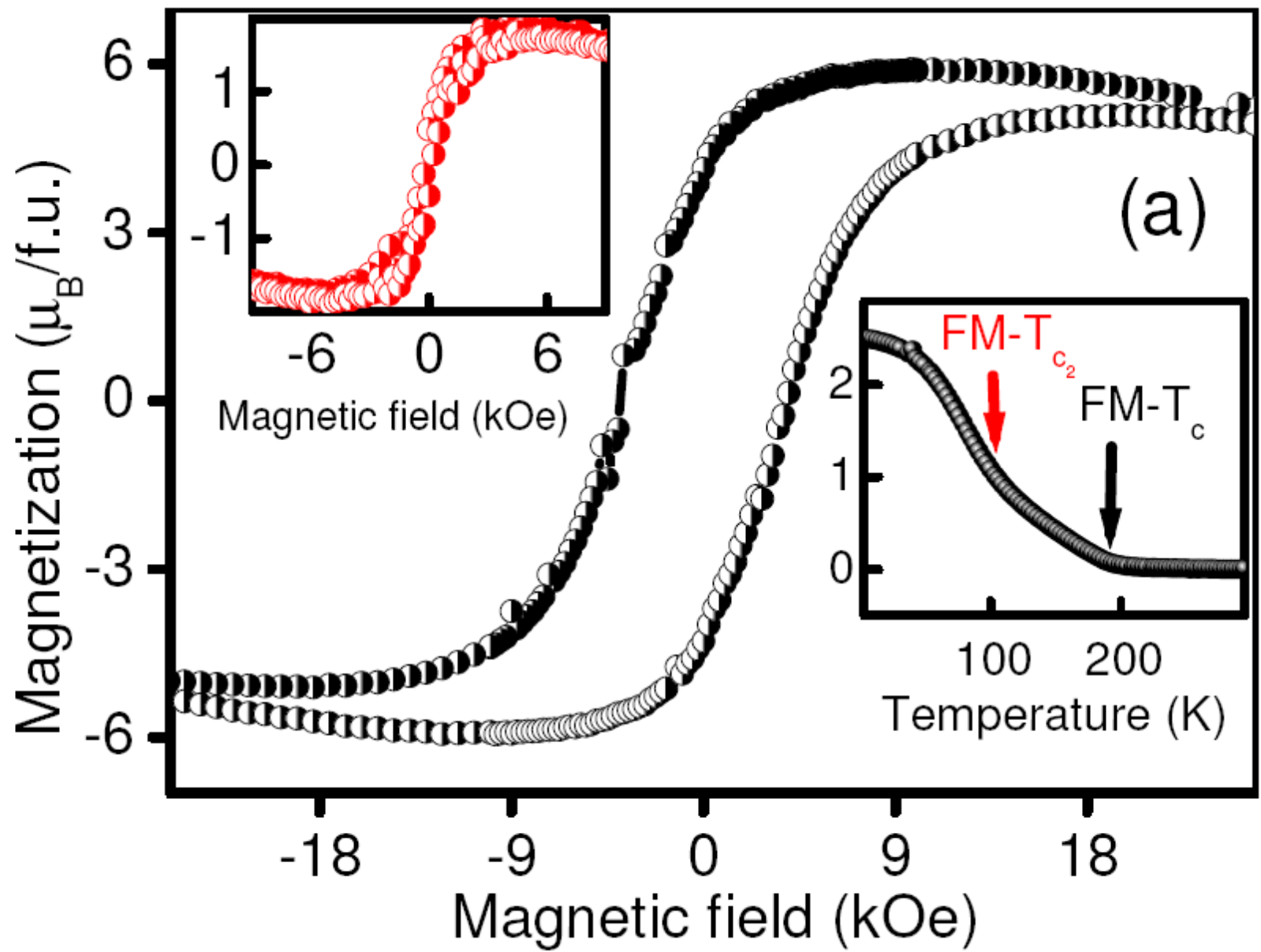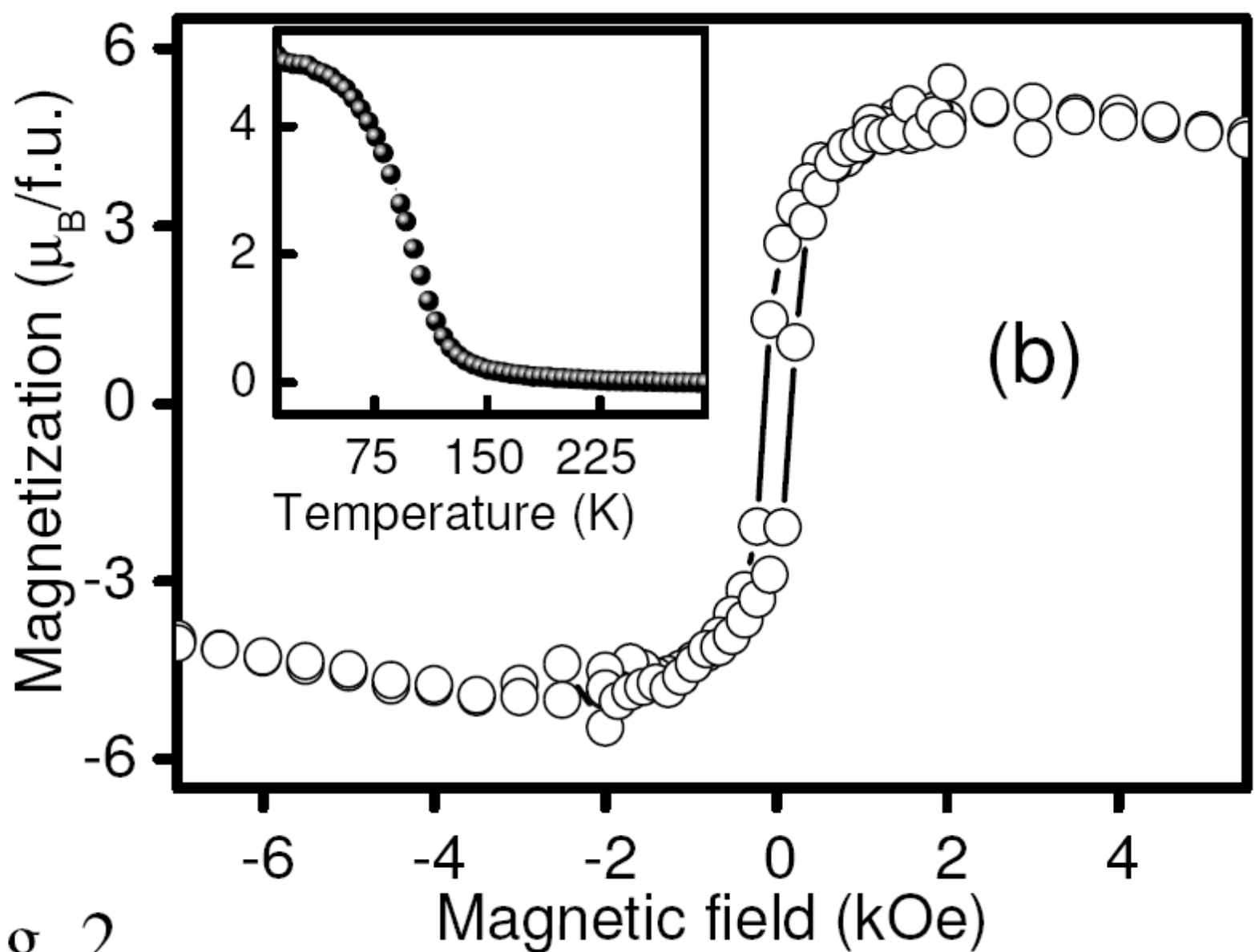

Fig. 2

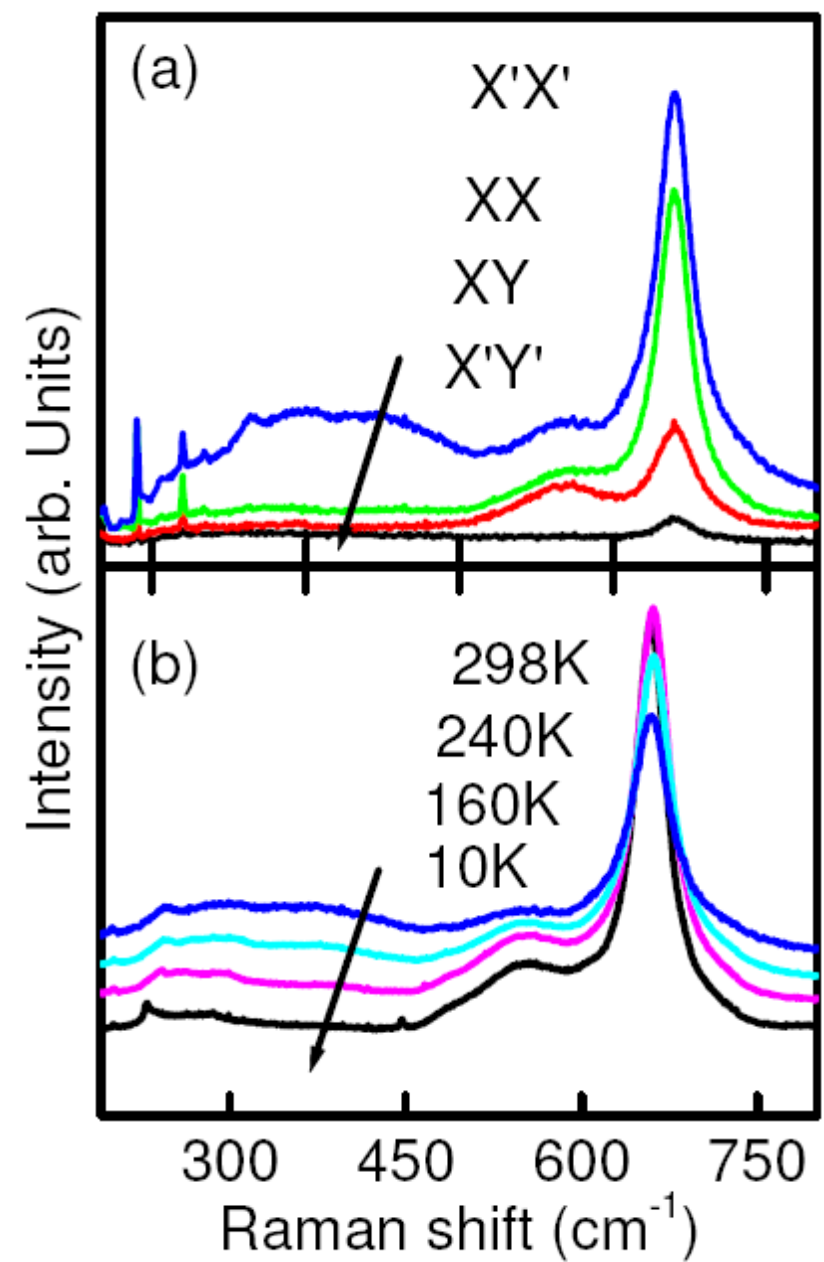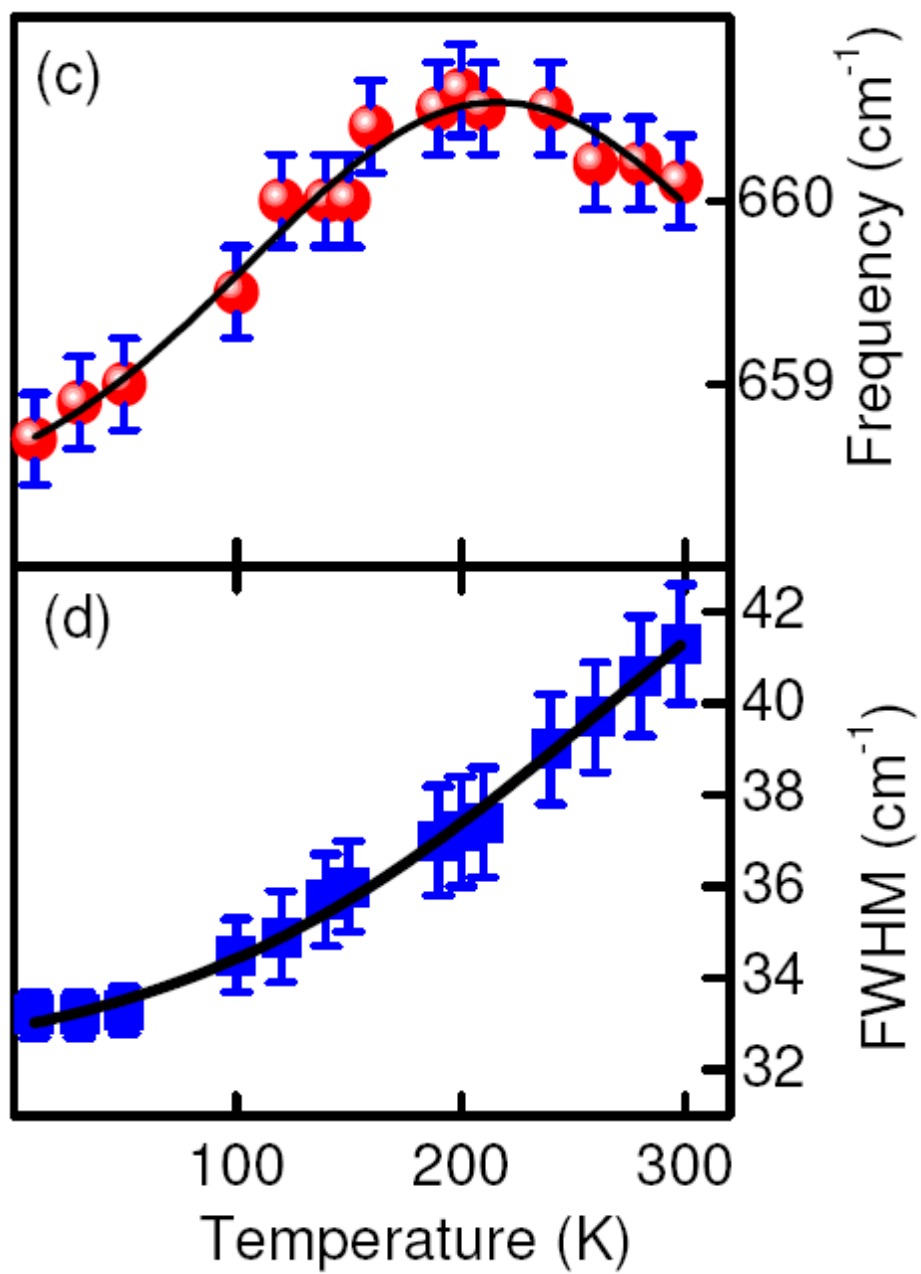

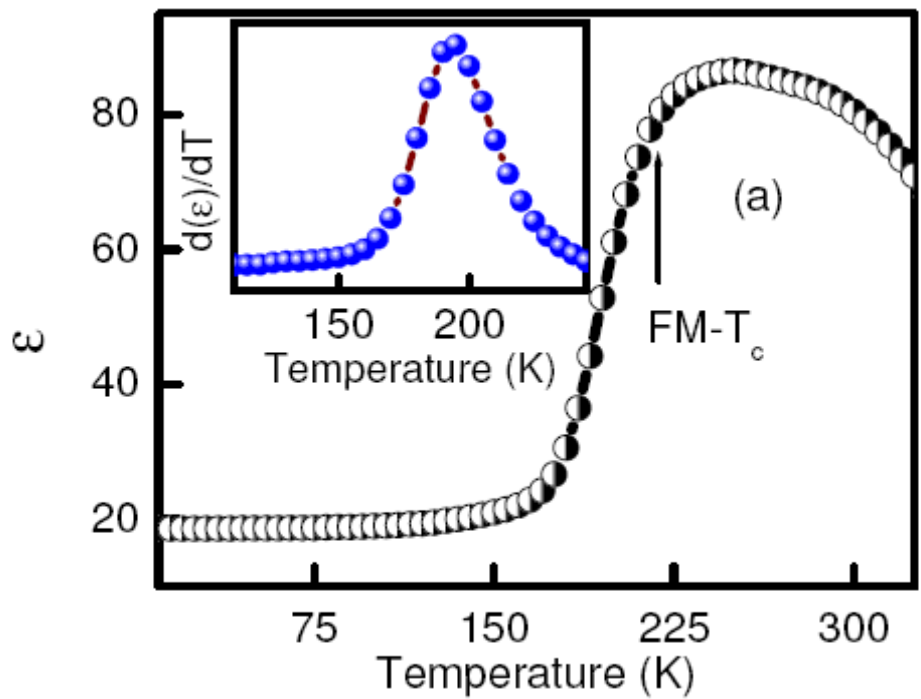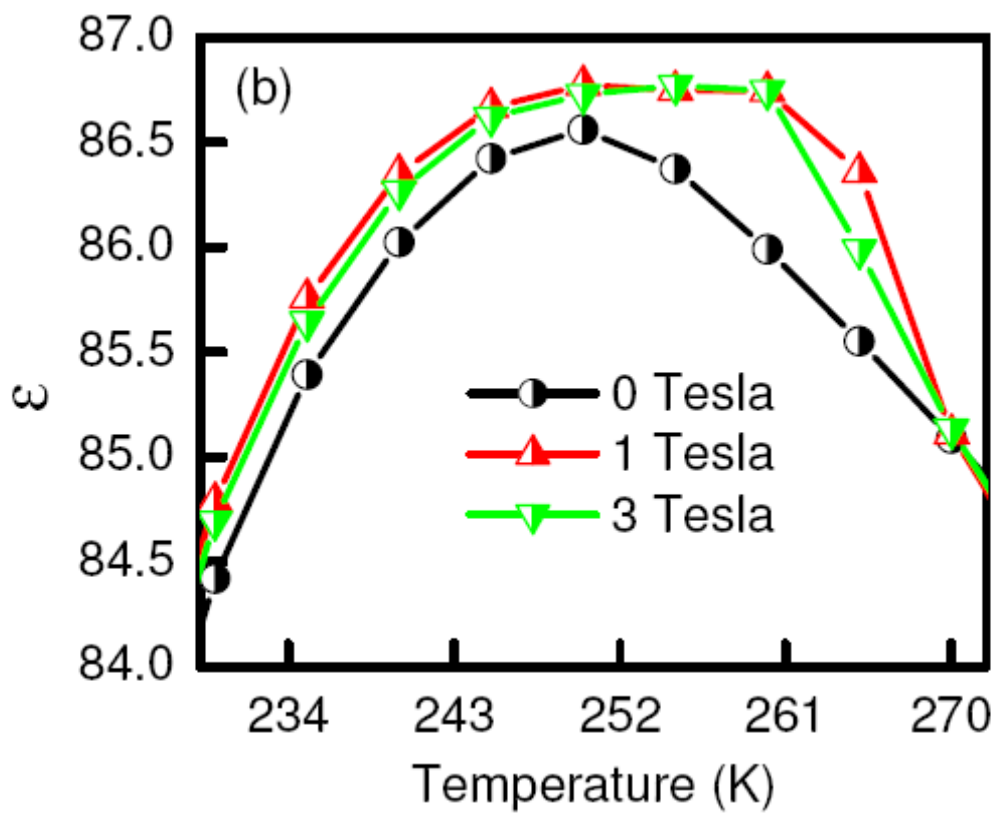